\documentstyle[agupp,epsf]{article}
\lefthead{DAUPHAS AND MARTY}
\righthead{THE LATE BOMBARDMENT OF EARTH}

\authoraddr{N. Dauphas and B. Marty,
Centre de Recherches P\'etrographiques et G\'eochimiques, CNRS UPR 2300, 15 rue Notre-Dame des Pauvres, BP 20, 54501 Vand\oe uvre-l\`es-Nancy Cedex France.
(e-mail: dauphas@crpg.cnrs-nancy.fr; bmarty@crpg.cnrs-nancy.fr)}

\slugcomment{Submitted to the {\it Journal of Geophysical Research}, 2001.}
\makeatletter
\let\reset@font\empty
\makeatother

\begin{document}

%% ------------------------------------------------------ %%
%
%%  TITLE
%
%% ------------------------------------------------------ %%

\title{Inference on the Nature and the Mass of Earth's Late Veneer\\ from Noble Metals and Gases}

	% Type the title of your manuscript between the
	% curly brackets in a \title command.  Capitalize
	% only acronyms, first letter of the first word,
	% first letter of proper nouns, first letter of
	% the first word after colons, and first letter
	% of the first word of a subtitle.  If the title
	% exceeds one line, break it so that the first line
	% is longer than the second line; break the title
	% before articles, prepositions, and conjunctions.
	% To break the title, type a double backslash where
	% you want the break to occur, as shown above.

%% ------------------------------------------------------ %%
%
%%  AUTHOR NAMES, AFFILIATIONS, and ALTERNATE AFFILIATIONS
%
%% ------------------------------------------------------ %%

\author{Nicolas Dauphas and Bernard Marty \altaffilmark{1}}
%
%\affil{Woods Hole Oceanographic Institution,
% Woods Hole, Massachusetts}
\affil{Centre de Recherches P\'etrographiques et G\'eochimiques, CNRS UPR 2300, Vand\oe uvre-l\`es-Nancy, France.}
%\author{J. Blythe and M. Chen\altaffilmark{1}}
%
%\affil{Observation Center for Prediction of
% Earthquakes and Volcanic Eruptions, Faculty
% of Science, Tohoku University, Sendai, Japan}

%\author{L. Song}
%
%\affil{Woods Hole Oceanographic Institution,
% Woods Hole, Massachusetts}

\altaffiltext{1}{Also at \'Ecole Nationale Sup\'erieure de G\'eologie, rue du doyen Marcel Roubault, B.P. 40, 54501 Vand\oe uvre-l\`es-Nancy Cedex France.}

\begin{abstract}
Noble metals and gases are very sensitive to the late accretion to the Earth of asteroids and comets. We present mass balance arguments based on these elements that indicate that $0.7-2.7\times 10^{22}$ kg of extraterrestrial bodies struck the Earth after core formation and that comets comprised less than $10^{-5}$ by mass of the impacting population. These results imply that the dynamics of asteroids and comets changed drastically with time and that biogenic elements and prebiotic molecules were not delivered to the Earth by comets but rather by carbonaceous asteroids.
\end{abstract}

\begin{article}
\section{Introduction}

The first recognizable microfossils are 3.5 Ga old and there is isotopic evidence that life was present on our planet before 3.8 Ga ago [\markcite {{\it Schopf,} 1993}; \markcite {{\it Mojzsis et al.,} 1996}]. The chain of events that led to the emergence of self-replicating organisms is poorly known but the late heavy bombardment of Earth by remnants of planetary formation might have played a vital role. The asteroidal and cometary battering that scarred the lunar surface had an ambivalent effect on the emergence and early evolution of life. On the one hand, large impacts caused global trauma of the biosphere but on the other hand, they contributed biogenic elements and prebiotic molecules. 

Approximately 65 Ma ago, a carbonaceous asteroid [\markcite {{\it Kyte,} 1998}] struck our planet at Chicxulub [\markcite {{\it Hildebrand et al.,} 1991}]. This impact probably caused the Cretaceous-Tertiary mass extinction [\markcite {{\it Alvarez et al.,} 1980}] and deposited intact organic molecules in sediments [\markcite {{\it Zhao and Bada,} 1989}]. Large impacts such as the K/T event imposed selection pressure on the biosphere  [\markcite {{\it Maher and Stevenson,} 1988}; \markcite {{\it Sleep et al.,} 1989}] and drove the evolutionary course from abiogenesis to the emergence of conciousness. Abiogenesis occured early in Earth's history [\markcite {{\it Schopf,} 1993}; \markcite {{\it Mojzsis et al.,} 1996}] while our planet was pummelled by an intense rain of carbonaceous asteroids and comets [\markcite {{\it Chyba,} 1990, 1991}; \markcite {{\it Dauphas et al.,} 2000}]. This late accreting veneer might have delivered biogenic elements and prebiotic molecules to the surface of Earth [\markcite {{\it Chyba and Sagan,} 1992}]. The lunar cratering record [\markcite {{\it Chyba,} 1990}], highly siderophile elements [\markcite {{\it Chyba,} 1991}], and water deuterium to protium ratios [\markcite {{\it Dauphas et al.,} 2000}] impose constraints on the mass of the late veneer ($M$ is the mass of asteroids and comets that struck the Earth in a late accretionary stage). What remains unanswered is whether the Earth was struck by stony or icy planetesimals ($\alpha$ is defined as the mass fraction of comets among impacting bodies). There is a stark contrast between the rare gas chemistry of asteroids and comets (\callout{Figure~1}), which raises the possibility that these elements can be used to address this question. In the present contribution, we show that the recent identification of argon in Hale-Bopp (C/1995 O1) [\markcite {{\it Stern et al.,} 2000}] places indeed stringent restrictions on the mass ($M$) and the nature ($\alpha$) of the late accreting veneer.

At present, the short-term accretion rate is dominated by sub-millimeter dust particles [\markcite {{\it Love and Brownlee,} 1993}]. Petrological and geochemical investigations conducted on micrometeorites indicate that the majority are carbonaceous [\markcite {{\it Kurat et al.,} 1994}; \markcite {{\it Engrand and Maurette,} 1998}]. Earth crossers, which dominate the long-term accretion rate, comprise a significant fraction of carbonaceous asteroids and comets [\markcite {{\it Shoemaker et al.,} 1990}]. Spectroscopic surveys of minor planets suggest that carbonaceous asteroids are overwhelmingly the most abundant type in outer regions of the main belt [\markcite {{\it Gradie et al.,} 1989}]. Also, the carbonaceous debris found in regolith breccias might represent a faithful time-integrated sampling of the asteroid belt population [\markcite {{\it Lipschutz et al.,} 1989}]. Additionally, dynamical modelling of early solar system evolution indicates that Earth's feeding zone extended with time to larger heliocentric distances, so that during the late accretionary stage our planet was impacted mostly by carbonaceous asteroids and comets [\markcite {{\it Morbidelli et al.,} 2000}]. For these reasons, we shall assume that the late veneer can be described as a binary mixture of carbonaceous asteroids and comets.

It is sometimes suggested that the osmium isotopic composition of the primitive upper mantle precludes direct filiation between mantle highly siderophile elements and carbonaceous chondrites [\markcite {{\it Meisel et al.,} 1996}; \markcite {{\it Meisel et al.,} 2001}; \markcite {{\it Walker et al.,} 2001}]. Core formation resulted in a fractionated noble metal pattern in the residual mantle. If a late carbonaceous veneer was mixed with this fractionated residue, then the osmium signature of the mantle is reconciliable with a carbonaceous late veneer [\markcite {{\it Dauphas et al.,} 2001}]. Note however that our conclusions do not rely heavily on this assumption.

\section{Noble Metals}

On Earth, geological processes such as erosion and tectonism obliterate impact structures so that the terrestrial cratering record does not exceed $\sim 2$ Ga [\markcite {{\it Grieve,} 1991}]. The earlier bombardment history of our planet can be inferred by examining the crater record of the Moon. The density of lunar craters increases with the age of the exposed surface in a manner indicative of an overall decline with time of the bombardment intensity [\markcite {{\it Shoemaker and Hackman,} 1962}; \markcite {{\it BVSP,} 1981}]. After scaling the lunar cratering record to Earth's gravitational cross section, it is estimated that $9.9\times 10^{20}-6.9\times 10^{23}$ kg of extraterrestrial bodies impacted our planet after the Moon formed [\markcite {{\it Chyba,} 1990}]. The lunar cratering record provides direct evidence for a decrease with time of the bombardment intensity but reliable quantification is difficult.

Additional evidence that asteroids and comets impacted our planet at a much higher rate in the Hadean than at present comes from water. The deuterium to protium ratio of the deep mantle [\markcite {{\it Deloule et al.,} 1991}] may be a remnant of the hydrogen isotopic composition of Earth forming planetesimals, which later evolved to the present terrestrial value as a result of the late accretion of volatile-rich matter [\markcite {{\it Kokubu et al.,} 1961}; \markcite {{\it Dauphas et al.,} 2000}]. If so, $4\times 10^{20}-2\times 10^{22}$ kg of asteroids and comets must have struck the Earth in a late accretionary stage [\markcite {{\it Dauphas et al.,} 2000}]. Note that this estimate depends on the mass fraction of comets among impacting bodies. Water deuterium to protium ratio provides a tighter estimate of the bombardment intensity than the lunar cratering record but the time interval over which integration holds is not very well constrained.

 The late bombardment of Earth by remnants of planetary formation left also its imprint in highly siderophile elements (Ru, Rh, Pd, Re, Os, Ir, Pt, and Au). When our planet differentiated, highly siderophile elements should have been partitioned almost completely into the core, leaving the mantle depleted and fractionated. This view is inconsistent with highly siderophile element abundances in mantle rocks [\markcite {{\it Kimura et al.,} 1974}; \markcite {{\it Jagoutz et al.,} 1979}]. A widely held opinion is that highly siderophile elements were delivered to the mantle by a late accreting veneer after metal/silicate differentiation [\markcite {{\it Morgan,} 1986}]. Thus, noble metals integrate the late bombardment of Earth over a time interval extending between core formation and present. Core formation is estimated from lead isotopes and ${\rm ^{182}Hf-^{182}W}$ systematics to have occurred approximately 50-100 Ma after collapse of the protosolar cloud [\markcite {{\it All\`egre et al.,} 1995}; \markcite {{\it Galer and Goldstein,} 1996}; \markcite {{\it Lee and Halliday,} 1995}; \markcite {{\it Halliday et al.,} 1996}]. Knowing the noble metal inventory of the silicate Earth, it is straightforward to calculate the mass of impacting bodies required to explain the observed highly siderophile element concentrations. We shall focus our discussion on osmium because its geochemistry is much better known than that of other noble metals. It has recently been suggested on the basis of the osmium isotopic composition of the mantle that there might be a concentration contrast for noble metals between the deep and the shallow mantle, the upper mantle being a factor of $3\pm 2$ richer than the lower mantle [\markcite {{\it Dauphas et al.,} 2001}]. The osmium concentration of the upper mantle is estimated to be $0.018\times 10^{-9}$ mol.g$^{-1}$ [\markcite {{\it McDonough and Sun,} 1995}], hence the osmium concentration of the deep mantle should be approximately $0.006\times 10^{-9}$ mol.g$^{-1}$ and there should be $3-7\times 10^{16}$ mol of osmium in the whole mantle. The osmium concentration of carbonaceous chondrites ranges from $2.6\times 10^{-9}$ to $4.3\times 10^{-9}$ mol.g$^{-1}$ and encompasses the range observed in other undifferentiated meteorite groups [\markcite {{\it Wasson and Kallemeyn,} 1988}]. Thus, the total mass of extraterrestrial matter accreted by the Earth after core formation must have been $0.7-2.7 \times 10^{22}$ kg. Examining the two extreme cases, highly siderophile elements are homogeneously distributed in the whole mantle and the deep mantle is devoid of noble metals, \markcite {{\it Chyba} [1991]} estimated that the mass accreted by the Earth after core formation must have been $1-4 \times 10^{22}$ kg, in complete agreement with the value we infer. These estimates rely on the assumption that the Earth was impacted by stony planetesimals. If the Earth was impacted by comets, then the total mass of extraterrestrial matter incident on Earth required to deliver the mantle inventory of noble metals would have been higher because the concentration of noble metals in comets must be lower than that of asteroids. Thus, the mass of extraterrestrial matter accreted by the Earth after core formation calculated on the basis of noble metals must increase along with the mass fraction of comets among impacting bodies (\callout{Figure~2}). Among the three approaches discussed so far, noble metals provide the tightest and most reliable estimate of $M$.

\section{Noble Gases}

We shall focus our discussion on argon because this element provides the most stringent constraints on the mass and the nature of the late accreting veneer. 

The net contribution of asteroids and comets to Earth's atmosphere is a balance between impact delivery and erosion [\markcite {{\it Chyba,} 1990}]. Detailed analytical and computational modeling of impact induced erosion indicate that impact events on Earth did not remove significant quantities of atmospheric gases [\markcite {{\it Newman et al.,} 1999}; \markcite {{\it Shuvalov and Artemieva,} 2000}]. It is sometimes suggested that rare gases were lost to space early in Earth's history as a result of an intense flux of extreme ultraviolet radiation from the young evolving sun [\markcite {{\it Zahnle and Walker,} 1982}]. The timing of such a phenomenon is not precisely known but it seems that Earth's atmosphere would have been retentive for argon $50-100$ Ma after collapse of the protosolar nebula [\markcite {{\it Hunten et al.,} 1987}]. As discussed previously, noble metals record the late bombardment of Earth beginning with core formation, which is thought to have occured $\sim 100$ Ma after formation of the solar system [\markcite {{\it All\`egre et al.,} 1995}; \markcite {{\it Galer and Goldstein,} 1996}; \markcite {{\it Lee and Halliday,} 1995}; \markcite {{\it Halliday et al.,} 1996}]. Thus, it is reasonable to assume that during the accretionary stage investigated in the present contribution (after metal/silicate differentiation) the atmosphere was retentive for argon. In addition, the ${\rm ^{38}Ar/^{36}Ar}$ ratio of the atmosphere [\markcite {{\it Ozima and Podosek,} 1983}] is unfractionated relative to that of the mantle [\markcite {{\it Kunz,} 1999}] and potential precursors [\markcite {{\it Mazor et al.,} 1970}], which casts doubts on the possibility that argon was ever lost to space.

If atmophiles were not lost to space by either impact erosion or hydrodynamic hydrogen escape, then the simplest mass balance requirement that the late veneer must meet is that it cannot deliver more than the terrestrial volatile inventory [\markcite {{\it Swindle and Kring,} 1997, 2001}]. This statement has strong implications on both the mass of asteroids and comets incident on Earth and the mass fraction of comets among impacting bodies. It is noteworthy that this mass balance holds even if Earth-forming planetesimals contributed a significant fraction of the terrestrial volatile inventory. The $^{36}$Ar inventory of the Earth provides a firm upper-limit on $M$, the mass of the late veneer.

Reliable estimates of the concentrations of argon in asteroids, comets, and the Earth are required before discussing further the mass and the nature of the late accreting veneer. Carbonaceous chondrites are the closest spectral analogs available in laboratory of the carbonaceous asteroids that impacted our planet during the Hadean eon [\markcite {{\it Gaffey et al.,} 1993}]. In these meteorites, the $^{36}$Ar concentration lies within  $2.3-6.2\times 10^{-11}$ mol.g$^{-1}$ [\markcite {{\it Mazor et al.,} 1970}].  

The appropriate cometary $^{36}$Ar content to use is a bit more difficult to determine, but can be constrained by studies of modern comets. Dynamical modeling of the early solar system evolution indicates that comets from the trans-Uranian region dominated the cometary flux in the neighborhood of Earth [\markcite {{\it Morbidelli et al.,} 2000}]. The ultimate source region of Oort cloud comets is uncertain but it might lie somewhere between the orbits of Uranus and Neptune [\markcite {{\it Fernandez and Ip,} 1981}]. In that case, the recent discovery of argon in Hale-Bopp [\markcite {{\it Stern et al.,} 2000}] is relevant to the late bombardment of Earth. The cometary $^{36}$Ar concentration is estimated to be $0.8-1.6\,\times 10^{-4}$ mol.g$^{-1}$ [\markcite {{\it Delsemme,} 1988}; \markcite {{\it Stern et al.,} 2000}]. Note that the presence of argon in solar proportions [\markcite {{\it Stern et al.,} 2000}] is consistent with condensation temperatures inferred from the low ortho-para ratios of water [\markcite {{\it Crovisier et al.,} 1997}].

The remaining piece of information required for calculation is the total $^{36}$Ar content of the Earth. For the sake of simplicity, we shall consider a layered mantle comprising the shallow mantle feeding mid-ocean ridges and the deep mantle feeding ocean islands, though we recognize that the true mantle structure may be considerably more complex [\markcite {{\it Tackley,} 2000}]. The shallow--deep mantle boundary lies somewhere between the transition zone and the D'' layer. The concentration of helium in the shallow mantle is calculated based on the observed flux at mid-ocean ridges [\markcite {{\it Craig et al.,} 1975}], the estimated rate of ocean crust formation [\markcite {{\it Crisp,} 1984}], and the inferred degree of partial melting [\markcite {{\it Klein and Langmuir,} 1987}]. The argon concentration in the shallow mantle is computed using isotopic and elemental ratios indexed to helium [\markcite {{\it Moreira et al.,} 1998}; \markcite {{\it Marty and Zimmermann,} 1999}]. Following the procedure outlined by \markcite {{\it All\`egre et al.} [1996]}, the radiogenic argon produced in Earth since 4.5 Ga by the decay of potassium is calculated [\markcite {{\it McDonough and Sun,} 1995}] and the amounts residing both in the shallow mantle and at the surface of Earth are then substracted in order to estimate the deep mantle $^{40}$Ar content. Using deep mantle ${\rm ^{40}Ar/^{36}Ar}$ ratios, it is then straightforward to calculate the deep mantle $^{36}$Ar content [\markcite {{\it Valbracht et al.,} 1997}; \markcite {{\it Marty et al.,} 1998}; \markcite {{\it Trieloff et al.,} 2000}]. Earth's core is thought to contain a negligible amount of argon [\markcite {{\it Matsuda et al.,} 1993}]. Uncertainties were propagated in the calculation by means of Monte-Carlo simulations [\markcite {{\it Anderson,} 1976}]. The $^{36}$Ar concentration of the bulk Earth is thus estimated to be $9.5-10.4\times 10^{-13}$ mol.g$^{-1}$. It is worthwhile to note that the argon inventory of Earth is not plagued by large uncertainty because most $^{36}$Ar resides in the atmosphere. Simply stated, approximately half of the $^{40}$Ar produced in Earth by the decay of potassium resides in the mantle [\markcite {{\it All\`egre et al.,} 1996}]. The ${\rm ^{40}Ar/^{36}Ar}$ of the silicate Earth is at least a factor of ten higher than that of the atmosphere [\markcite {{\it Valbracht et al.,} 1997}; \markcite {{\it Marty et al.,} 1998}; \markcite {{\it Trieloff et al.,} 2000}], so that at most one tenth of the terrestrial $^{36}$Ar inventory resides in the mantle.

The atmophile inventories compiled in \callout{Table~1} were computed in a similar manner to that detailed in the case of argon.

  The mass balance requirement on argon dictates that only a limited region of the $\alpha - M$ space is admissible (hatched region labelled Noble Gases in Figure~2). Outside this region, too much ${\rm ^{36}Ar}$ would have been delivered to the Earth. As discussed earlier, noble metals provide independent constraints on $M$ (hatched region labelled Noble Metals). Combining these two admissible locii, the mass of asteroids and comets incident on Earth after core formation must have been $0.7-2.7\times 10^{22}$ kg and comets must have represented less than $10^{-5}$ by mass of the impacting population (cross hatched region). Note that  \markcite {{\it Swindle and Kring} [2001]} also reached the conclusion that comets represented a minor fraction of Earth's late veneer by similar lines of argument. This conclusion is not very sensitive to the exact concentration of Ar in comets and will probably remain valid as more data become available.

\section{Perspectives}

At present, comets comprise a significant fraction of Earth's impacting population, at least on the order of a few percent [\markcite {{\it Shoemaker et al.,} 1990}]. On the contrary, our results indicate that comets represented a negligibly small fraction of the extraterrestrial bodies that impacted our planet in a late accretionary stage. Thus, the dynamics of asteroids and comets must have changed drastically with time, which calls for theoretical confirmation. 

 The mass fraction of comets among impacting bodies is so small that they could not have contributed significant quantities of biogenic elements and prebiotic molecules to Earth. In constrast, carbonaceous asteroids [\markcite {{\it Kerridge,} 1985}] could have delivered a significant fraction of the present surface inventory [\markcite {{\it Schlesinger,} 1997}] of hydrogen, carbon, and nitrogen. These biogenic elements might have been delivered in the form of prebiotic molecules and might also have been reprocessed by impact shockwaves into organic molecules [\markcite {{\it Chyba and Sagan,} 1992}].

The finding that the mass of asteroids and comets incident on Earth after core formation was $0.7-2.7\times 10^{22}$ kg and that comets represented less than $10^{-5}$ by mass of the impacting population has important implications for both early solar system dynamics and the emergence and early evolution of life on Earth. Further investigations must be conducted on comets, specifically on their rare gas chemistry, in order to obtain a clearer picture of the origin of planetary atmospheres and Earth's biosphere.

\acknowledgments
We thank V.V. Shuvalov, L. Reisberg, and F. Robert for stimulating discussions. This research has made use of NASA's Astrophysics Data System Abstract Service. This is CRPG contribution xxxx.

\end{article}

%% ------------------------------------------------------ %%
%
%%  FIGURE CAPTIONS
%
%% ------------------------------------------------------ %%

   % Set figure captions between the curly brackets in the
   % \caption command.  Each figure caption must have a
   % \begin{figure} before and an \end{figure} after,
   % as shown.  Several captions may be printed per page,
   % as long as there is sufficient room to cut between
   % the captions.
   %
   % It may be necessary to use \clearpage commands
   % between a long series of captions, since a large
   % number of captions can cause LaTeX's memory buffer
   % to overload and crash.
   %
   % AGU asks authors to submit two sets of captions;
   % one set for single column and one set for double
   % column figures.  You must input the first set of
   % captions, but AGU's LaTeX style files will
   % automatically generate the second set of captions.
   % Both sets of captions will print at the correct
   % single- and double-column widths for your journal
   % (these widths change depending on the document style
   % you choose).
   %
   % The \figurewidth commands do not affect the second
   % set of figure captions.  The second set of captions
   % will always mirror the figure numbers of the first
   % set of captions.
   %
   % The wide figure captions will likely cause
   % "overfull hbox" error messages to appear when
   % you LaTeX your file: please ignore these.
   %
   % Be aware that figure width and figure numbering
   % commands may affect following plate captions,
   % and vice versa.
   %
   % Caption lines may be broken by using a \protect
   % command in front of a \linebreak command:
   % \protect\linebreak (this will maintain justification).

\begin{figure}[p]
        \begin{center}
        \leavevmode
        \epsfbox{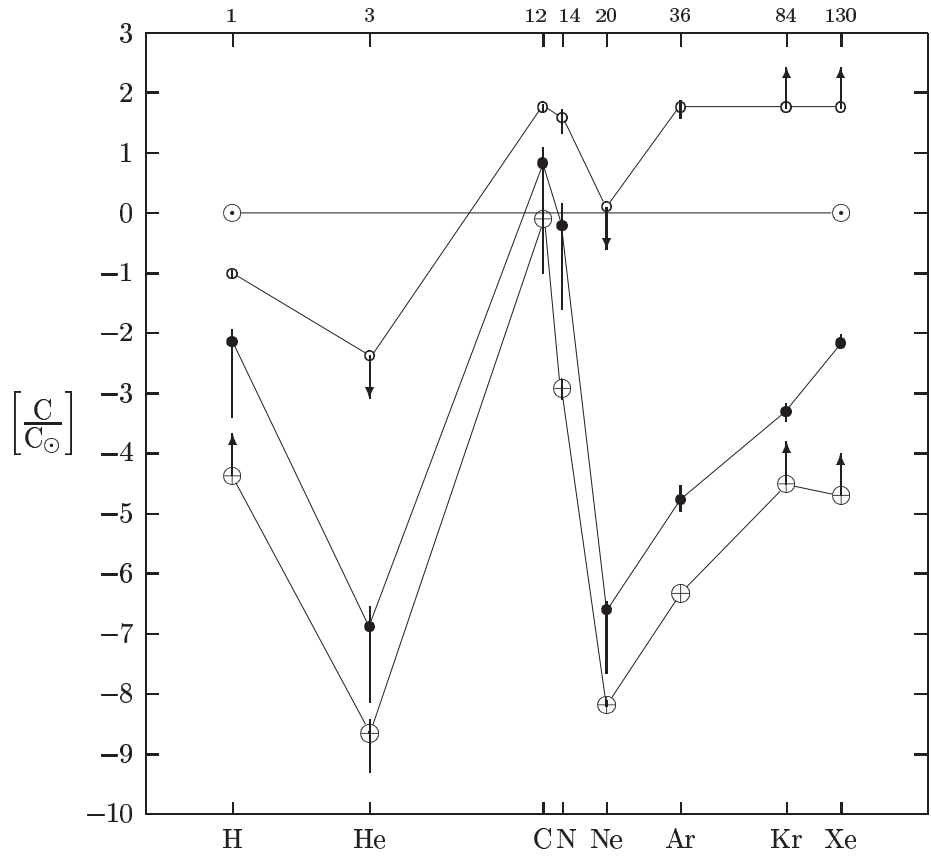}
        \end{center}
        \caption{Concentrations of hydrogen, helium, carbon, nitrogen, neon, argon, krypton, and xenon in asteroids ($\bullet$), comets ($\circ$), and Earth ($\earth$) are normalized to cosmic abundances of the elements ($\odot$). Values and references are compiled in Table 1.}
        \end{figure}

\begin{figure}[p]
        \begin{center}
        \leavevmode
        \epsfbox{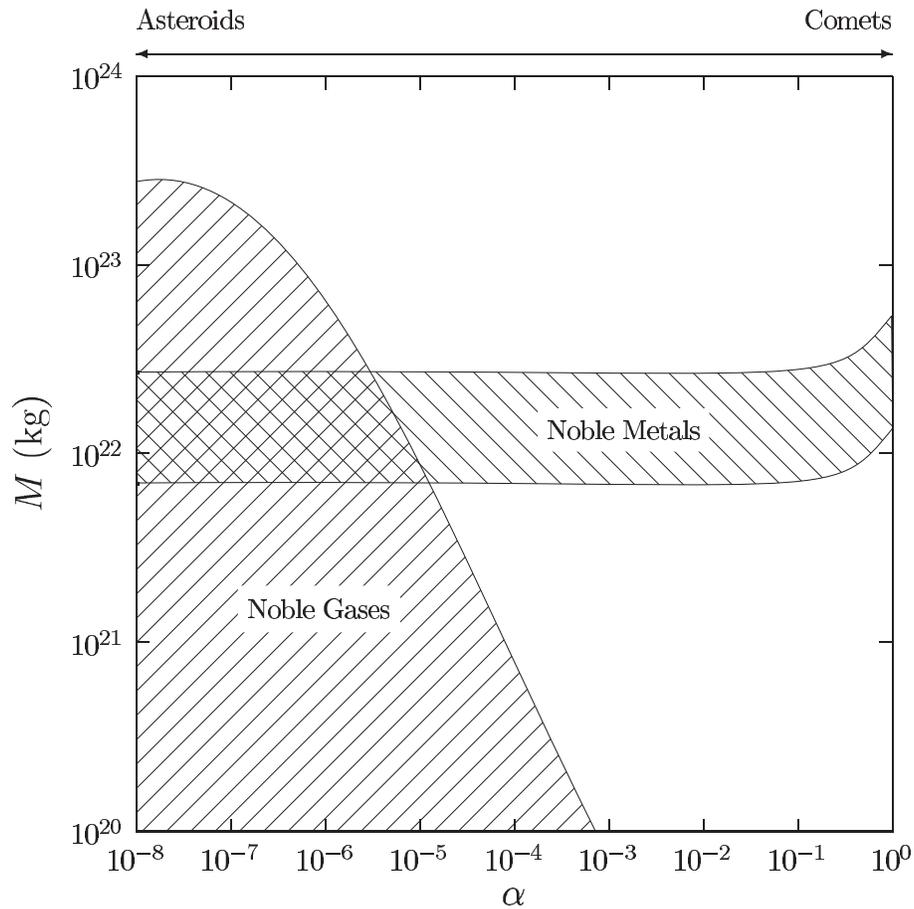}
        \end{center}
\caption{The mass of asteroids and comets that struck the Earth after core formation ($M$) is reported as a function of the mass fraction of comets among impacting bodies ($\alpha$). The admissible locii from the points of view of noble metals and gases are indicated as hatched regions. The admissible region for both noble metals and gases is the intersection of these two fields (cross hatched). Thus, we estimate that the mass of asteroids and comets incident on Earth after core formation was $0.7-2.7\times 10^{22}$ kg and that comets represented less than $10^{-5}$ by mass of the impacting population (see text for details).}
\end{figure}

\newpage

\begin{planotable}{lrcccccccc}
\tablewidth{40pc}
\tablecaption{Atmophile Inventories}
\tablenum{1}
\tablehead{& &$^1$H & $^3$He & $^{12}$C & $^{14}$N & $^{20}$Ne & $^{36}$Ar & $^{84}$Kr & $^{130}$Xe}
\tablenotetext{\null}{Concentrations are expressed in mol.g$^{-1}$ and power of ten multipliers are in parentheses. Cosmic -- \markcite {{\it Anders and Grevesse,} 1989}. Asteroids -- \markcite {{\it Kerridge,} 1985}; \markcite {{\it Mazor et al.,} 1970}. Comets -- \markcite {{\it Bar-Nun and Owen,} 1998}; \markcite {{\it Delsemme,} 1988}; \markcite {{\it Jessberger et al.,} 1988}; \markcite {{\it Krasnopolsky et al.,} 1997}; \markcite {{\it Stern et al.,} 1992, 2000}. Earth -- 
\markcite {{\it All\`egre et al.,} 1996};
\markcite {{\it Craig et al.,} 1975};
\markcite {{\it Crisp,} 1984}; 
\markcite {{\it Dauphas and Marty,} 1999};
\markcite {{\it Klein and Langmuir,} 1987};
\markcite {{\it McDonough and Sun,} 1995}; 
\markcite {{\it Marty and Tolstikhin,} 1998}; 
\markcite {{\it Marty et al.,} 1998}; 
\markcite {{\it Marty and Zimmermann,} 1999};
\markcite {{\it Michael,} 1988}; 
\markcite {{\it Moreira et al.,} 1998}; 
\markcite {{\it Ozima and Podosek,} 1983}; 
\markcite {{\it Schlesinger,} 1997};  
\markcite {{\it Trieloff et al.,} 2000}; 
\markcite {{\it Valbracht et al.,} 1997}. 
}
\startdata
Cosmic & & 7.0(-1) & 9.6(-6) & 2.5(-4) &  7.8(-5) & 8.0(-5) & 2.1(-6) & 6.5(-10) & 5.2(-12) \nl
\nl
Asteroids & {\Large \{} & {\Large $^{2.8(-4)}_{8.1(-3)}$} & {\Large $^{7.0(-14)}_{2.7(-12)}$} & {\Large $^{2.9(-4)}_{3.1(-3)}$} & {\Large $^{2.0(-6)}_{1.1(-4)}$} & {\Large $^{1.8(-12)}_{2.8(-11)}$} & {\Large $^{2.3(-11)}_{6.2(-11)}$} & {\Large $^{2.2(-13)}_{4.5(-13)}$} & {\Large $^{2.9(-14)}_{5.0(-14)}$}\nl
\nl
Comets & {\Large \{} & {\Large $^{5.7(-2)}_{7.7(-2)}$} & $<4(-8)$ & {\Large $^{1.2(-2)}_{1.6(-2)}$} & {\Large $^{1.7(-3)}_{4.1(-3)}$} & $<1(-4)$ & {\Large $^{0.8(-4)}_{1.6(-4)}$} & $>3(-8)$ & $>3(-10)$\nl
\nl
Earth & {\Large \{} & $>3(-5)$ & {\Large $^{4.3(-15)}_{3.7(-14)}$} & {\Large $^{3.5(-5)}_{3.5(-4)}$} & {\Large $^{7.2(-8)}_{1.3(-7)}$} & {\Large $^{4.8(-13)}_{5.5(-13)}$} & {\Large $^{9.5(-13)}_{1.0(-12)}$} & $>2(-14)$ & $>1(-16)$ 
\end{planotable}

\newpage

%\thispagestyle{empty}
%\begin{figure}
%\begin{center}
%\input{./figure0.tex}
%\end{center}
%\label{fig:Figure 1}
%\end{figure}
\end{document}